\documentstyle[12pt]{article}
\newcommand{\RR}{$\rm l\!R\;\;$}

\newcommand{\nut}{
\begin{picture}(11,5)(0,0)
\put(5,-10){$\tilde{}$}
\put(0,0){\it N}
\end{picture}}

\newcommand{\fu}{
\begin{picture}(9,5)(0,0)
\put(2.5,10){\tiny 1}
\put(0,0){F}
\end{picture}}

\newcommand{\fd}{
\begin{picture}(9,5)(0,0)
\put(2.5,10){\tiny 2}
\put(0,0){F}
\end{picture}}

\newcommand{\bu}{
\begin{picture}(9,5)(0,0)
\put(2.5,12){\tiny 1}
\put(0,0){$\tilde{{\rm B}}$}
\end{picture}}

\newcommand{\bd}{
\begin{picture}(9,5)(0,0)
\put(2.5,12){\tiny 2}
\put(0,0){$\tilde{{\rm B}}$}
\end{picture}}

\newcommand{\au}{
\begin{picture}(9,5)(0,0)
\put(3,10){\tiny 1}
\put(0,0){A}
\end{picture}}

\newcommand{\ad}{
\begin{picture}(9,5)(0,0)
\put(2.5,10){\tiny 2}
\put(0,0){A}
\end{picture}}

\newcommand{\cu}{
\begin{picture}(11,5)(0,0)
\thicklines
\put(0,9){\line(1,-2){5}}
\put(0,9){\line(1,0){10}}
\thinlines
\put(5,-1){\line(1,2){5}}
\put(3.7,3.7){\tiny 1}
\end{picture}}

\newcommand{\cd}{
\begin{picture}(11,5)(0,0)
\thicklines
\put(0,9){\line(1,-2){5}}
\put(0,9){\line(1,0){10}}
\thinlines
\put(5,-1){\line(1,2){5}}
\put(3.7,3.7){\tiny 2}
\end{picture}}

\begin{document}
\baselineskip=24pt plus 0.2pt minus 0.2pt
\lineskip=22pt plus 0.2pt minus 0.2pt
\begin{center}
 \Large
A Real Polynomial Formulation of General\\
Relativity in terms of Connections\\

\vspace*{0.35in}

\large

J.\ Fernando\ Barbero\ G. $^{\ast, \dag}$
\vspace*{0.25in}

\normalsize

$^{\ast}$Center for Gravitational Physics and Geometry\\
104 Davey Lab, Pennsylvania State University,\\
University Park, PA 16802\\
U.S.A.\\

$^{\dag}$Instituto de Matem\'aticas y F\'{\i}sica Fundamental, \\
C.S.I.C.\\
Serrano 119--123, 28006 Madrid, Spain
\\

\vspace{.5in}
November 11, 1993\\
\vspace{1.5in}
ABSTRACT

\end{center}
I show in this letter that it is possible to construct a Hamiltonian
description for Lorentzian General Relativity in terms of two real
$SO(3)$ connections. The constraints are simple polynomials in
the basic variables. The present framework gives us a new formulation of
General Relativity that keeps some of the interesting features of
the Ashtekar formulation without the complications associated
with the complex character of the latter.


\pagebreak

\setcounter{page}{1}

Using a somewhat  conservative approach in which both General
Relativity and the general quantization program are taken as the starting
point, A. Ashtekar has developed a new description for Gravity that has
some features that may, at the end, lead to a successful quantization of
the theory \cite{Ash}\cite{Ashlibro}. A key element of his approach is
changing  the emphasis  from geometrodynamics to connection dynamics.
In fact it is not only the simplicity of the constraints in Ashtekar's
Hamiltonian formulation that makes it possible to advance in the
quantization program but also the availability of geometrical objects that
are absent in the geometrodynamical description. This fact is at the
root of the successful introduction  of the loop variables  by Rovelli and
Smolin \cite{R}. They have provided a very appealing picture of the
structure of space at the Planck scale and made it possible to find
solutions to all the constraints in the Hamiltonian formulation of the theory.

In spite of all the deep new insights that have been gained, it must be said
that the program is not complete in its present form and we cannot claim
yet that gravity has been successfully quantized. There are many technical
issues that must be addressed (and many conceptual questions
too). One of them, the intrinsically complex nature of the Ashtekar
variables, will be the subject of this letter. The fact that the Ashtekar
connection must be genuinely complex is something that can be seen both
at the levels of the Hamiltonian and the Lagrangian descriptions. In the
Hamiltonian formalism an imaginary unit must be introduced in the
canonical transformation that leads from the usual geometrodynamical
phase space coordinates to the Ashtekar variables. It is
necessary if one wants to eliminate some terms involving derivatives of
the triads that would complicate the final form of the Hamiltonian
constraint. In the Lagrangian description (using the
Samuel-Jacobson-Smolin action \cite{Sam}\cite{JaS})  use is made of
self-dual connections that are, again, complex in the Lorentzian case. The
real theory is recovered by imposing "reality conditions" on the fields.
Although these conditions may prove to be useful in order to obtain the
inner product of the theory, in practice they are difficult to implement. It
is, in my opinion, desirable to have manifestly real formulations
of General Relativity that avoid these problems while keeping the
simplicity of the Ashtekar approach (or, at least, a significant part of it).

The main result presented in this letter is to show that it is possible to
describe 3+1 dimensional gravity in a phase space spanned by two real
$SO(3)$ connections (much in the spirit of \cite{fer}) with constraints
that are low order polynomials in the phase space variables both for
Euclidean and Lorentzian General Relativity. Although this approach is close
to the Ashtekar point of view of using connections as the basic objects to
describe the gravitational field, the geometric nature of the fields
involved is different, and thus it may allow us to find new sets of
elementary variables for the quantization of gravity that are not obvious in
the previous formulations.

A point that I want to discuss before proceeding further is the meaning of
polynomiality and its relationship with the "reality" of a formulation. As
has been stressed by several authors (see \cite{tate} and references
therein) the geometrodynamical constraints can always be cast in a
polynomial form by introducing powers of the determinant of the 3-metric
as global factors. The issue is not really whether the constraints are
polynomial but how simple their polynomial expressions are. If in the
Ashtekar formulation we do not introduce an imaginary unit in the
canonical transformation that brings us from the Lorentzian ADM phase
space to the new one but work, instead, with real fields we find a real
formulation in terms of "Ashtekar-like" variables. The problem is that
proceeding in this way the final Hamiltonian constraint has a complicated
expression and high density weight (if one wants it to be polynomial).
This makes very difficult the passage to the quantum theory in which we
must impose the quantum version of the constraints as conditions on the
wave functionals. Some of the advantages of working with the Ashtekar
variables are then lost. However, the fact that the basic variables are
still the connection-densitized triad pair may still allow us to use, for
example, the loop variable approach when attempting the quantization of
the theory and get some interesting results. It must be emphasized, also,
that the really important issue seems to be finding a simple way to write
the {\it quantum} constraints, and so it is conceivable that a somewhat
complicated set of elementary variables could do the trick and give a
simple quantum theory. Even if a formulation does not have simple
constraints, the geometrical nature of the basic variables may suggest a
set of elementary variables that simplifies the quantum theory. This, in
itself, is a motivation to describe General Relativity using different
sets of basic variables.

In the following I will further exploit the ideas introduced in \cite{fer} to
describe 3+1 complex General Relativity in a phase space coordinatized by
two  complex  $SO(3)$ connections. I will start by giving an argument that
shows that it may be possible to find an appealing polynomial formulation
for Lorentzian General Relativity  and then give the full construction.

The phase space introduced in \cite{fer} to describe complex General
Relativity is coordinatized by two complex$SO(3)$ connections
$\au_{a}^{i}$ and $\ad_{a}^{i}$. I
introduce here the notation that will be used throughout the letter.
Tangent space indices and $SO(3)$ indices are represented by Latin
letters from the beginning and the end of the alphabet respectively, and run
from 1 to 3. The space-time manifold is restricted to have the form
${\cal M}=$\RR$ \!\!\!\!\times\Sigma$ where $\Sigma$ is a compact
3-manifold with no boundary. I introduce also the objects:
$e_{ai}\equiv \ad_{ai}-\au_{ai}$, $\bu_{i}^{a}\equiv\tilde{\eta}^{abc}
\fu_{bci}$, $\bd_{i}^{a}\equiv\tilde{\eta}^{abc}\fd_{bci}$, $\det e\equiv
\tilde{{\rm e}}$ and $\tilde{b}_{i}^{a}\equiv \bd^{a}_{i}- \bu_{i}^{a}$, where
$\tilde{\eta}^{abc}$ is the 3-dimensional Levi-Civita tensor density,
$\epsilon_{ijk}$ is the internal Levi-Civita tensor and $\fu_{ab}^{i}$,
$\fd_{ab}^{i}$ are the curvatures of $\au_{a}^{i}$ and $\ad_{a}^{i}$
($F_{ab}^{i}\equiv 2\partial_{[a} A_{b]}^{i}+\epsilon^{i}_{\;\;jk}A_{a}^{j}
A_{b}^{k}$). I represent the density weights by the usual convention
of using tildes above and below the fields.

The symplectic structure in this model is given by \cite{fer}
\begin{equation}
\Omega=2\kappa\int_{\Sigma}d^{3}x\;\tilde{\eta}^{abc}\epsilon_{ijk}
\left[\au_{a}^{i}(x)-\ad_{a}^{i}(x)\right]d\au_{b}^{j}(x)\wedge d\ad_{c}^{k}
(x)
\label{1}
\end{equation}
\noindent where $\kappa=i$ and $\kappa=1$ for Lorentzian and Euclidean
gravity, respectively). The constraints are \cite{fer}
\begin{eqnarray}
\epsilon_{i}^{\;\;jk}e_{aj}\bu_{k}^{a}=0 & &\nonumber\\
\epsilon_{i}^{\;\;jk}e_{aj}\bd_{k}^{a}=0 & &\label{2}\\
e_{a}^{k}\bu_{k}^{a}=0 & &\nonumber
\end{eqnarray}
(the Lagrange multipliers for the Gauss law and the diffeomorphism
constraint must be taken as purely imaginary in the Lorentzian case). The
scalar constraint has now density weight +1 because this formulation is
well defined only when the $e_{a}^{i}$ are non degenerate (a condition
that may be traced back to the non-degeneracy of the symplectic 2-form
\cite{fer}). We can then drop a factor $\tilde{{\rm e}}$ that appears in
the Hamiltonian constraint.

Following the arguments given by Hojman, Kucha\v{r} and Teitelboim
\cite{HKT} it is straightforward to see that the 3-metric is just
$q_{ab}=e_{a}^{i}e_{bi}$. Introducing now the Hamiltonian constraint
functional \begin{equation}
H(\nut)=\int_{\Sigma}d^{3}x\;\nut(x)\tilde{{\rm e}}\;e_{a}^{i}\bu^{a}_{i}
\label{3}
\end{equation}
\noindent it is possible to compute the Poisson bracket of $H(\nut)$
and $q_{ab}$ to get the extrinsic curvature in terms of the two connections
\begin{equation}
\left\{H(\nut),
q_{ab}\right\}=\frac{1}{2}\kappa\left[-\nut{\tilde{\rm e}}q_{ab}+\frac{1}{2}
\nut\left(q_{ab}e_{c}^{k}\tilde{b}^{c}_{k}-2q_{e(a}e_{b)}^{k}
\tilde{b}^{e}_{k}\right)\right]\equiv -2NK_{ab}
\label{4}
\end{equation}
\noindent where I
define $N\equiv\nut\tilde{{\rm e}}$. One possibility now is to impose
reality conditions. One must demand that both the 3-metric $q_{ab}$ and
the  extrinsic curvature $K_{ab}$ given by (\ref{4}) are real (the presence
of $\kappa$ will give non trivial conditions for  the Lorentzian case). We
can, however, realize that the expression  $\tilde{q}[K^{ab}K_{ab}-K^{2}]$
($\tilde{q} \equiv \det q_{ab}$) that appears in the usual Hamiltonian
constraint of geometrodynamics is now a simple polynomial in
$e_{a}^{i}$ and $\tilde{b}^{a}_{i}$. This, together with the fact that
$e_{a}^{i}$ and $\tilde{b}^{a}_{i}$ are canonically conjugate (as can
be seen from the Poisson brackets derived from (\ref{1})) suggests that it
may be possible to find a change of coordinates from the
geometrodynamical phase space (or rather the triad-extrinsic curvature
version of it) to a phase space coordinatized by two real connections in
which the constraints are simple polynomials. The reason why one expects
this to happen is  because, even in the Lorentzian case, the dangerous
terms quadratic in the extrinsic curvatures have simple polynomial
expressions as deduced from (\ref{4}).

I give now the complete construction of the new formulation.
The starting point will be the geometrodynamical description of Lorentzian
General Relativity in terms of  the triad and
the extrinsic curvature. The phase space is $\Gamma(\tilde{E}, K)$
coordinatized by $\tilde{E}_{i}^{a}$ and $K_{ai}$ with the symplectic
structure
\begin{equation}
\Omega=\int_{\Sigma}d^{3}x\;dK_{a}^{i}\wedge d\tilde{E}_{i}^{a}
\label{5}
\end{equation}
\noindent and the constraints
\begin{eqnarray}
2q^{-1/2}\tilde{E}^{b}_{[i}\tilde{E}^{a}_{j]}K_{a}^{i}K_{b}^{j}+\zeta
q^{1/2}R=0 & &\nonumber\\
{\cal D}_{b}(K_{a}^{i}\tilde{E}^{b}_{i}-K_{c}^{i}\tilde{E}^{c}_{i}
\delta_{a}^{b})=0 & &\label{6}\\
\epsilon_{ijk}K_{a}^{j}\tilde{E}^{ak}=0 & &\nonumber
\end{eqnarray}
\noindent Here ${\cal D}_{a}$ is the torsion-free covariant derivative,
compatible with $e_{ai}$  that acts both on internal and spatial indices
(${\cal D}_{a}e_{bi}\equiv\partial_{a}e_{bi}+\epsilon_{i}^{\;\;jk}\Gamma_{aj}
e_{bk}-\Gamma_{ab}^{\;\;\;\;c}e_{ci}=0$) and $\zeta=-1$ or $\zeta=1$ for
Lorentzian and Euclidean General Relativity, respectively.
The constraints (\ref{6}) generate time evolution, spatial diffeomorphisms
and $SO(3)$ gauge transformations \cite{abj}
The previous symplectic structure gives the Poisson brackets
\begin{eqnarray}
& & \left\{K_{a}^{i}(x), K_{b}^{j}(y)\right\}=0\nonumber\\
& & \left\{\tilde{E}^{a}_{i}(x), \tilde{E}^{b}_{j}(y)\right\}=0\label{7}\\
& & \left\{\tilde{E}^{a}_{i}(x),
K_{b}^{j}(y)\right\}=\delta_{b}^{a}\delta_{i}^{j}\delta^{3}(x,y)\nonumber
\end{eqnarray}

Let us introduce now a change of coordinates from the geometrodynamical
phase space to a new one coordinatized by two {\it real} $SO(3)$
connections $\au_{a}^{i}$, $\ad_{a}^{i}$
\begin{eqnarray}
& &
\tilde{E}_{i}^{a}=\tilde{\eta}^{abc}\epsilon_{ijk}e_{b}^{j}e_{c}^{k}\label{8}\\
& & K_{a}^{i}=\frac{1}{4\tilde{{\rm
e}}}\left[e_{a}^{i}e_{b}^{j}-2e_{a}^{j}e_{b}^{i}\right]\tilde{b}^{b}_{j}
\label{9}
\end{eqnarray}

\noindent It is straightforward to check that the Jacobian of the previous
transformation is well defined and different from zero if and only if
$\tilde{{\rm e}}\neq 0$. Substituting now (\ref{8}) and (\ref{9}) in (\ref{5})
we conclude that the symplectic structure can be written in terms of the
two connections as in (\ref{1}) with $\kappa=1$. The Poisson brackets in
terms of these new variables are
\begin{eqnarray}
& & \left\{\au_{a}^{i}(x), \au_{b}^{j}(y)\right\}=0\nonumber\\
& & \left\{\ad_{a}^{i}(x), \ad_{b}^{j}(y)\right\}=0\label{10}\\
& & \left\{\au_{a}^{i}(x), \ad_{b}^{j}(y)\right\}=\frac{1}{4\tilde{{\rm e}}}
\left[e_{a}^{i}e_{b}^{j}-2e_{a}^{j}e_{b}^{i}\right]\delta^{3}(x,y)\nonumber
\end{eqnarray}

They coincide with the result obtained in \cite{fer} for the
Husain-Kucha\v{r} model. We must see now how the constraints (\ref{6})
are written in terms of the two connections. As far as the Gauss law and
vector constraints are concerned, we expect to find the result already
obtained in \cite{fer} and given by the first two expressions in (\ref{2}).
Indeed these constraints give us just the kinematical symmetries of the
theory. For the Gauss law we find that it is translated into the condition
\begin{equation}
\epsilon_{ijk}e_{a}^{j}\tilde{b}^{ak}=0\label{11}
\end{equation}
\noindent which is equivalent to
\begin{equation}
\cu_{b}\bd^{b}_{i}+\cd_{b}\bu^{b}_{i}=0
\label{11a}
\end{equation}
The diffeomorphism constraint ${\cal
D}_{b}(K_{a}^{i}\tilde{E}^{b}_{i}-K_{c}^{i}\tilde{E}^{c}_{i}\delta_{a}^{b})
+\epsilon_{ijk}\Gamma_{a}^{i}K_{b}^{j}\tilde{E}^{bk}=0$ becomes
now
\begin{equation}
\au_{a}^{i}\cu_{b}\bd^{b}_{i}+\ad_{a}^{i}\cd_{b}\bu^{b}_{i}=0
\label{12}
\end{equation}
In the last expressions $\cu$ and $\cd$ denote the covariant derivatives
built out of the connections $\au_{a}^{i}$ and $\ad_{a}^{i}$ and defined by
$\cu_{a}\lambda_{b}=\partial_{a}\lambda_{b}+\epsilon_{i}^{\;\;jk}\au_{aj}
\lambda_{k}$ (and the analogous expression for $\cd$).
Using the Bianchi identities it is easy to show now that, when
$\tilde{{\rm e}}\neq 0$, Eqs. (\ref{11}) and (\ref{12}) are equivalent to
\begin{eqnarray}
& & \epsilon_{i}^{\;\;jk}e_{aj}\bu_{k}^{a}=0\nonumber\\
& & \epsilon_{i}^{\;\;jk}e_{aj}\bd_{k}^{a}=0\label{13}
\end{eqnarray}

\noindent The generating functionals of $SO(3)$ gauge transformations and
diffeomorphisms are given by (see \cite{fer})
\begin{eqnarray}
& & G(N)=-\int_{\Sigma}d^{3}x
N^{i}\left[\cu_{b}\bd^{b}_{i}+\cd_{b}\bu^{b}_{i}\right]\nonumber\\
& & D(\vec{N})=\int_{\Sigma}d^{3}x N^{a}\left[\ad_{a}^{i}\cd_{b}\bu^{b}_{i}
+\au_{a}^{i}\cu_{b}\bd^{b}_{i}\right]\label{14}
\end{eqnarray}

Let us concentrate now on writing the Hamiltonian constraint in terms of
two connections. As a preliminary step it is useful to point out that the
general solution to the equation
\begin{equation}
2\tilde{\eta}^{abc}\epsilon_{ijk}e_{b}^{j}A_{c}^{k}=\tilde{H}^{a}_{i}
\label{15}
\end{equation}
\noindent (where $A_{c}^{k}$ is the unknown) in terms of $e_{a}^{i}$
and $\tilde{H}^{a}_{i}$ is equal to
\begin{equation}
A_{a}^{i}=\frac{1}{4\tilde{{\rm
e}}}\left[e_{a}^{i}e_{b}^{j}-2e_{a}^{j}e_{b}^{i}\right]\tilde{H}^{b}_{j}
\label{16}
\end{equation}
The $SO(3)$ connection $\Gamma_{a}^{i}$ compatible with $e_{ai}$ is
given by a particular case of (\ref{16})
\begin{equation}
\Gamma_{a}^{i}=-\frac{1}{2\tilde{{\rm
e}}}\left[e_{a}^{i}e_{b}^{j}-2e_{a}^{j}e_{b}^{i}\right]\tilde{\eta}^{bcd}
\partial_{c}e_{dj}
\label{17}
\end{equation}
Taking into account that $e_{a}^{i}\equiv\ad_{a}^{i}-\au_{a}^{i}$ we see
that
\begin{equation}
\tilde{b}^{a}_{i}-2\tilde{\eta}^{abc}\partial_{b}e_{ci}=\tilde{\eta}^{abc}
\epsilon_{ijk}e_{b}^{j}(\au_{c}^{k}+\ad_{c}^{k})
\label{18}
\end{equation}
\noindent and solving (\ref{18}) for the real $SO(3)$ connection
$A_{a}^{i}=1/2(\au_{a}^{i}+\ad_{a}^{i})$ we find, using Eqs.
(\ref{9}) and (\ref{17}) that
\begin{equation}
A_{a}^{i}=\frac{1}{4\tilde{{\rm
e}}}[e_{a}^{i}e_{b}^{j}-2e_{a}^{j}e_{b}^{i}][\tilde{b}^{b}_{j}-2\eta^{bcd}
\partial_{c}e_{dj}]=K_{a}^{i}+\Gamma_{a}^{i}
\label{19}
\end{equation}
\noindent The previous expression gives
$\Gamma_{a}^{i}(\au,\ad)=A_{a}^{i}(\au,\ad)-K_{a}^{i}(\au,\ad)$. We
can use it to simplify the computation of the scalar
curvature term that appears in the geometrodynamical scalar
constraint in terms of the two curvatures. In fact, we can follow the
procedure used by Ashtekar to write
the Hamiltonian constraint for both Euclidean and Lorentzian General
Relativity. We must keep in mind that the final expressions should be
written in terms of $\au_{a}^{i}$ and $\ad_{a}^{i}$.

In the Euclidean case we have
\begin{equation}
\tilde{{\rm e}}^{2}\epsilon^{ijk}e^{a}_{i}e^{b}_{j}[F_{abk}(\au,\ad)-2{\cal
D}_{a}K_{bk}(\au,\ad)]=0
\label{20}
\end{equation}
\noindent where
$F_{abk}(\au,\ad)=\frac{1}{2}\fu_{abk}+\frac{1}{2}\fd_{abk}-\frac{1}{4}
\epsilon_{k}^{\;\;lm}e_{al}e_{bm}$. For Lorentzian General Relativity we
find
\begin{equation}
\tilde{{\rm e}}^{2}\epsilon^{ijk}e^{a}_{i}e^{b}_{j}[F_{abk}(\au,\ad)+
2\epsilon_{klm}K_{a}^{l}(\au,\ad)K_{b}^{m}(\au,\ad)
-2{\cal D}_{a}K_{bk}(\au,\ad)]=0
\label{21}
\end{equation}
\noindent Taking into account that ${\cal D}_{a}$ is compatible with
$e_{ai}$, we see that the term involving ${\cal D}_{a}K_{bk}$ is proportional
to the Gauss law and hence we can remove it. In terms of $\au_{a}^{i}$ and
$\ad_{a}^{i}$ the Hamiltonian constraints for Euclidean and Lorentzian
General relativity become respectively
\begin{eqnarray}
& & \tilde{{\rm e}}\left[e_{c}^{k}\bu^{c}_{k}+e_{c}^{k}\bd_{k}^{c}-3
\tilde{{\rm e}}\right]=0\label{22}\\
& & \tilde{{\rm e}}\left[e_{c}^{k}\bu^{c}_{k}+e_{c}^{k}\bd_{k}^{c}-
3\tilde{{\rm e}}
\right]-\frac{1}{2}\left[e_{b}^{k}e_{c}^{l}-2e_{b}^{l}e_{c}^{k}\right]
\tilde{b}^{b}_{k}\tilde{b}^{c}_{l}=0\label{23}
\end{eqnarray}
\noindent Although the term quadratic in $\tilde{b}^{a}_{i}$ in (\ref{23})
makes this expression more complicated than its Euclidean counterpart, it
is still a low order polynomial in the basic variables and has density weight
+2, just as the Hamiltonian constraint in the Ashtekar formulation. It is
possible to further simplify (\ref{22}) and (\ref{23}) by using the following
canonical transformation
\begin{eqnarray}
& & \au_{a}^{i}\rightarrow \au_{a}^{i}+\alpha e_{a}^{i}\label{24}\\
& & \ad_{a}^{i}\rightarrow \ad_{a}^{i}+\alpha e_{a}^{i}\label{25}
\end{eqnarray}
\noindent where $\alpha$ is a real constant. Under the action of (\ref{24})
and (\ref{25}), $\bu^{a}_{i}$, $\bd^{a}_{i}$, $\tilde{b}^{a}_{i}$ transform as
\begin{eqnarray}
& &
\bu^{a}_{i}\rightarrow(1-\alpha)\bu^{a}_{i}+\alpha\bd^{a}_{i}+
\alpha(\alpha-1)\tilde{\eta}^{abc}\epsilon_{ijk}e_{b}^{j}e_{c}^{k}\label{26}\\
& &
\bd^{a}_{i}\rightarrow(1+\alpha)\bu^{a}_{i}-\alpha\bd^{a}_{i}+
\alpha(\alpha+1)\tilde{\eta}^{abc}\epsilon_{ijk}e_{b}^{j}e_{c}^{k}\label{27}\\
& & \tilde{b}^{a}_{i}\rightarrow
\tilde{b}^{a}_{i}+2\alpha\tilde{\eta}^{abc}\epsilon_{ijk}e_{b}^{j}e_{c}^{k}
\label{28}
\end{eqnarray}
The Gauss law and the diffeomorphism constraint are invariant under the
action of the previous canonical transformation, whereas the Hamiltonian
constraints (\ref{22}) and (\ref{23}) transform respectively into
(remember that $\tilde{{\rm e}}\neq0$)
\begin{eqnarray}
& & \tilde{{\rm e}}\left[(1\!-\!2\alpha)e_{c}^{k}\bu^{c}_{k}+(1\!+\!2\alpha)
e_{c}^{k}\bd^{c}_{k}+ 3(4\alpha^2\!-\!1)\tilde{{\rm e}}\right]=0\label{29}\\
& & \tilde{{\rm e}}\left[(1\!+\!2\alpha) e_{c}^{k}\bu^{c}_{k}+(1\!-\!2\alpha)
e_{c}^{k}\bd^{c}_{k}+ 3(12\alpha^2\!-\!1)\tilde{{\rm  e}}\right]\!-\!
\frac{1}{2}(e_{b}^{k}e_{c}^{l}-2e_{b}^{l}e_{c}^{k})\tilde{b}^{b}_{k}
\tilde{b}^{c}_{l}=0\label{30}
\end{eqnarray}
By choosing now some $\alpha$ we can simplify the previous expressions.
If we take, for example, $\alpha=-1/2$ we get
\begin{eqnarray}
& & e_{c}^{k}\bu^{c}_{k}=0\label{31}\\
& & \tilde{{\rm e}}\left[e_{c}^{k}\bd^{c}_{k}+3\tilde{{\rm
e}}\right]-\frac{1}{4}\left[e_{b}^{k}e_{c}^{l}-2e_{b}^{l}e_{c}^{k}\right]
\tilde{b}^{b}_{k}\tilde{b}^{c}_{l}=0\label{32}
\end{eqnarray}
for Euclidean and Lorentzian General Relativity respectively.
The first expression reproduces the result found in \cite{fer} for complex
General Relativity, as one would expect, due to the triviality of the reality
conditions in the Euclidean case. The second one gives a polynomial
Hamiltonian constraint for Lorentzian General Relativity.

Several comments are in order at this point. The formulation described
above in terms of $SO(3)$ connections is polynomial of low order in the
basic variables and real by construction. It avoids, then, the introduction of
reality
conditions. This may prove to be an advantage of this formalism. In fact,
reality conditions are difficult to deal with even in the pure gravity case.
It is not known, for example, how to implement them if one uses the loop
variables to quantize the theory.
It must be said, nevertheless, that reality conditions may also be a
useful tool. They can be used, for example, to select the scalar product of the
theory \cite{Ashlibro} as can be shown in
several non-trivial examples like electromagnetism or linearized gravity.

Of course, we must
decide if the simplification brought about by the real character of the
theory  compensates for the complications associated with both the non
trivial symplectic structure and the presence of terms quadratic in the
curvatures in the Hamiltonian constraint. As far as the symplectic
structure is concerned, it must be said that, although it is not trivial, it
can
be found in some familiar examples such as the two connection
formulation of the Husain-Kucha\v{r} model \cite{HuKu}\cite{fer}. This
kind of symplectic structure may be actually be a common feature of
theories formulated in terms of two connections.  The structure of
(\ref{32}) is not as simple as in the Euclidean formulation but has some
nice features. In a sense, it is somewhat in between the Hamiltonian
constraints in the ADM and the Ashtekar formulations; actually the last
term in (\ref{32}) exactly corresponds to the term quadratic  in the
extrinsic curvatures that appears in the ADM scalar constraint whereas
the first term is close in form to the Euclidean Hamiltonian constraint, and
thus, to the direct translation of the Hamiltonian constraint of the
Ashtekar formulation to the two connection phase space.

The fact that $\au_{a}^{i}$ and $\ad_{a}^{i}$ are not canonically
conjugate makes it difficult to talk about a canonical quantization of the
theory. This
means that the passage to the quantum theory is not straightforward in
this  formulation. One really needs to find a set of elementary variables
suitable for the task. Although this may not be easy it must be said that,
in spite of the complications associated with the symplectic structure, it is
possible to find pairs of canonically conjugate variables that are not
obvious in the Ashtekar formalism. One example of this is given by
$e_{a}^{i}$ and $\tilde{b}^{a}_{i}$.

Another feature that makes the previous framework appealing is the way
degenerate metrics can be taken care of. They are simply excluded by the
requirement that the symplectic structure be non-degenerate (or
equivalently by the condition  that the coordinate transformation
introduced above is well defined). It may be that degenerate metrics
convey interesting physical information and it may be even possible to
satisfactorily deal with them (as it happens in 2+1 dimensions when one
uses Witten's  formulation \cite{Witt}). Nevertheless they are known
to be a possible source of trouble as has been  emphasized by Smolin
\cite{smo} and clearly shown by Varadarajan \cite{madh} with his example
of a spherically symmetric solution to  all the constraints of General
Relativity in 3+1 dimensions that is regular everywhere, degenerate  in
some regions, and has arbitrary negative energy. My opinion is that it is
certainly reassuring  to have a consistent way of dealing with degenerate
metrics.

The ideas presented above strongly suggest that it may be  possible to
find a real action for Lorentzian General Relativity in terms of two real
connections. The  inclusion of matter in this action could provide an
explicit  realization of the idea suggested by Ashtekar \cite{Ashlibro}  of
unifying all the interactions as a consequence of the fact that gravity can
be described in terms of connections.  Although some work in  this
direction has already been done \cite{Peld}\cite{cdj}, it may be worth
looking at this problem from the.perspective of the real, two connection
formulation presented above.

{\bf Acknowledgements} I want to thank A. Ashtekar, G. Mena,
J. Mour\~ao, P. Peld\'an and M. Varadarajan for interesting discussions and
comments and I acknowledge financial support provided by the Consejo
Superior de Investigaciones Cient\'{\i}ficas (Spain). I am also grateful to
Guillermo Mena and Madhavan Varadarajan for their careful reading of the
manuscript.

\end{document}